\documentclass[footinbib, twocolumn, english, aps, prl, showpacs, superscriptaddress, floats, amsmath, amssymb, floatfix]{revtex4}

\usepackage{color}
\usepackage{amsmath}
\usepackage{graphicx}
\usepackage{graphics}
\usepackage[pdf]{pstricks}
\usepackage{pdftexcmds}
\usepackage{amssymb}
\usepackage{esint}
\usepackage{comment}
\usepackage{physics}
\usepackage{bm}
\usepackage{lipsum}                

\def \c {\hat{c}}
\def \cd {\hat{c}^{\dagger}}

\def \d {\hat{d}}
\def \dd {\hat{d}^{\dagger}}

\newcommand{\Dp}{\mathbf{D}}
\newcommand{\Sp}{\mathbf{S}}
\newcommand{\Se}{\mathbf{s}}

\def\Rv {{\bf R}}
\def\rv {{\bf r}}
\def\kv {{\bf k}}

\def \bea {\begin{eqnarray}}
\def \eea {\end{eqnarray}}
\def \c {\hat{c}}
\def \cd {\hat{c}^{\dagger}}

\allowdisplaybreaks

\begin{document}

\title{Topological phase transition driven by magnetic field and topological
    Hall effect in an antiferromagnetic skyrmion lattice}

\author{M. Tom\'e}
\affiliation{Instituto de Física de Líquidos y Sistemas Biológicos (IFLYSIB), UNLP-CONICET, Facultad de Ciencias Exactas, La Plata, Argentina}
\affiliation{Departamento de Física, Facultad de Ciencias Exactas, Universidad Nacional de La Plata, La Plata, Argentina}
\author{H.D. Rosales}
\affiliation{Instituto de Física de Líquidos y Sistemas Biológicos (IFLYSIB), UNLP-CONICET, Facultad de Ciencias Exactas, La Plata, Argentina}
\affiliation{Departamento de Física, Facultad de Ciencias Exactas, Universidad Nacional de La Plata, La Plata, Argentina}
\affiliation{Departamento de Cs. B\'asicas, Facultad de Ingenier\'ia,
Universidad Nacional de La Plata, C.C. 67, 1900 La Plata, Argentina}
\date{\today}

\begin{abstract}
The topological Hall effect (THE), given by a composite of electric and topologically non-trivial spin texture is commonly observed in magnetic skyrmion crystals. Here we present a study of the THE of electrons coupled to antiferromagnetic Skyrmion lattices (AF-SkX). We show that, in the strong Hund coupling  limit, topologically non-trivial phases emerge at specific fillings. Interestingly, at low filling  an external field controlling the magnetic texture, drives the system from a conventional insulator  phase to a phase exhibiting THE.
Such behavior suggests the occurrence of a topological transition which is confirmed by a closing of the bulk-gap that is followed by its reopening, appearing simultaneously with a  single pair of helical edge states. This transition is further verified by the calculation of the the Chern numbers and Berry curvature. We also compute a variety of observables in order to quantify the THE, namely: Hall conductivity and the orbital magnetization of electrons moving in the AF-SkX texture. 
\end{abstract}

\maketitle

{\it Introduction.--} Magnetic skyrmions, a kind of topologically protected solitons,  are nanometer size spin textures that have shown to be of great practical interest due both to the fundamental properties  and its promising potential in spintronics based applications\cite{nagaosa2013topological,fert2013skyrmions}. They have been evidenced experimentally in a wide variety of materials including chiral magnets such as MnSi \cite{muhlbauer2009skyrmion,jonietz2010spin}, FeGe \cite{yu2011near}, Fe$_x$Co$_{1-x}$Si \cite{yu2010real} and $\beta$-Mn-type Co-Zn-Mn \cite{tokunaga2015new}; and insulator materials as  Cu2OSeO3\cite{seki2012observation,seki2012magnetoelectric}. 
In most cases, periodic arrays of skyrmions are stabilized by the competition of strong ferromagnetic exchange interactions, external magnetic field  and the antisymmetric Dzyaloshinskii-Moriya (DM) interaction \cite{bogdanov1989thermodynamically,bogdanov1994thermodynamically,bocdanov1994properties,roessler2006spontaneous}. However, in recent years, theoretical studies have suggested that skyrmions might be also stabilized in antiferromagnets where frustration helps to stabilize antiferromagnetic skyrmions crystals (AF-SkX) consisting of multiple interpenetrated ferromagnetic Skyrmion lattices \cite{rosales2015three,barker2016static,zhang2016antiferromagnetic,fujita2017ultrafast,jin2016dynamics,osorio2017composite,osorio2019stability,osorio2019skyrmions,villalba2019field}. Recently, in the spinel MnSc$_2$S$_4$ the first realization of an AF-SkX (fractional) like-structure has been discovered, where the planes (triangular lattices) perpendicular to the field direction host interpenetrated fractional ferromagnetic Skyrmion lattices \cite{ShangGao2020Nat}.

When conduction electrons are coupled to the local magnetic background, they accumulate a Berry phase as they travel through skyrmions spin configuration which acts as a local effective magnetic field leading to topological Hall effect (THE) \cite{THE1,THE2,THE3,THE4,THE5,THE6,ndiaye2017topological,rosales2019frustrated}. Most of the previous studies on THE have focused on systems consisting of ferromagnetic SkX showing unconventional behaviour that emerges as a consequence of the non-trivial smooth magnetic texture \cite{gobel2017unconventional,gobel2017signatures,gobel2018family}. 
However, a not so desired feature of skyrmions hosted in ferromagnets is that they exhibit an inevitable topology effect, namely, the skyrmion Hall effect \cite{jiang2017direct}. In this phenomena, the magnetic skyrmions do not move collinear to the current flow direction, but acquire a transverse motion due to the appearance of a topological Magnus force acting upon the non-zero topological charge. To avoid this disadvantage, theoretical studies have suggested that the skyrmion Hall effect can be suppressed by utilizing the counterpart of the ferromagnetic skyrmions: the antiferromagnetic skyrmions, which are also topologically protected but without showing the skyrmion Hall effect \cite{gobel2020beyond}. 
In this context, electrons coupled  to  antiferromagnetic skyrmion lattices AF-SkX have  been less explored, even knowing that the multiple sublattice structure can induce interesting magnetic phenomena \cite{gobel2017antiferromagnetic}. 
Therefore a question that arises naturally is whether conduction electrons coupled to this kind of skyrmion lattices could present exotic new physics.

\begin{figure}[thb]
\includegraphics[width=0.8\columnwidth]{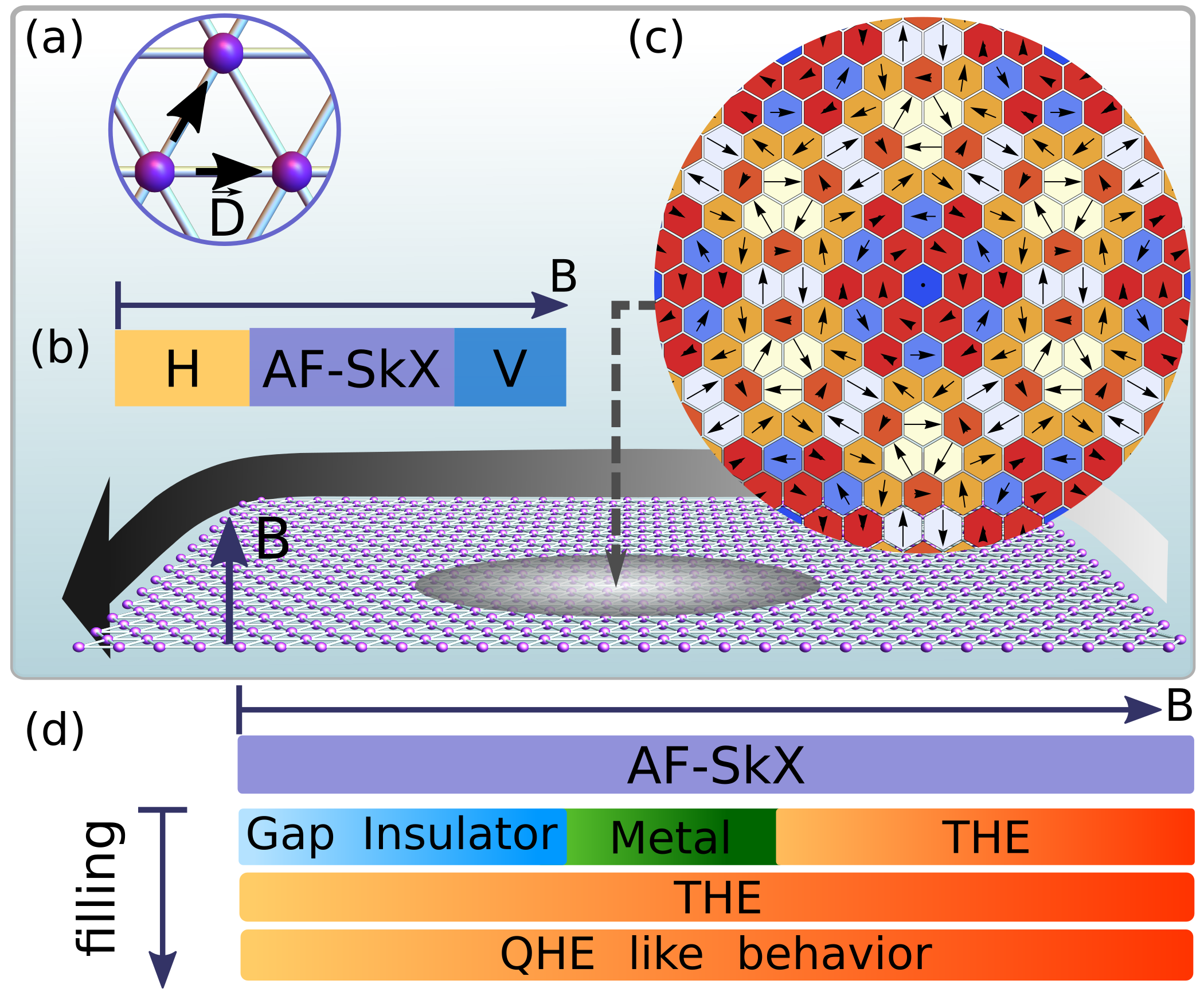}
\caption{(Color online) (a) DM  interactions ${\bf D}$ (black arrows). (b) Low temperature phase diagram of the magnetic Hamiltonian in Eq. (\ref{eq:Hm}) including Helical (H), AF-SkX and vortex like (V) phases. (c)  Portion of an AF-SkX magnetic background considered in this paper for $D/J=0.5$ and $B/J=4.6$. (d) Schematic phase diagram  of the electronic Hamiltonian in Eq. (\ref{eq:hamiltinian_eff}). At low filling, the magnetic field induces a sequence of topological transitions: band insulator $\to$ metal $\to$ THE; at high filling the system's  behaviour can be connected with Integer Quantum Hall Effect.} 
\label{fig:lattice}
\end{figure}

This letter address this relevant question on the THE of electrons in topologically non-trivial AF-SkX textures on the triangular lattice. By means of extensive Monte-Carlo simulations and exact diagonalization for the  magnetic and fermionic sector respectively, we find that in the strong Hund coupling limit, at low filling it is possible to control the THE and its protected edge states by tuning the external magnetic field. This phenomena is confirmed by a closing and further reopening of the bulk-gap and the transition from trivial to non-trivial Chern number. As a complement, we have computed the orbital magnetization, which is directly related to Berry-phase effect, showing surprisingly opposite signs at both sides of the topological transition. This property enables a ``switch'' on/off of the THE and orbital magnetization by an external magnetic field.

{\it Model and Methods.--} In this work we consider a tight-binding model on a triangular lattice where the interaction of electrons with an AF-SkX  texture is described by the Hamiltonian,
\bea
H&=&-\sum_{\langle \rv,\rv'\rangle,\sigma}t_{\rv\rv'}\,(\cd_{\rv\sigma}\c_{\rv',\sigma}+\text{h.c.})-J_h\sum_{\rv,\mu\nu}\Sp_{\rv}\cdot \Se_{\rv}\qquad
\label{eq:hamiltonian_tb}
\eea

\noindent where $\c_{\rv\sigma}$ ($\cd_{\rv\sigma}$) is the creation (annihilation) operator at the  site $\rv$ with spin ($\sigma=\uparrow,\downarrow$), $t_{\rv\rv'}$ is the hopping between nearest-neighbor sites, $J_H$ is the Hund's coupling strength between the electron spin $\Se_{\rv}=\frac{1}{2}\cd_{\rv,\mu}\vec{\sigma}^{\mu\nu}\c_{\rv,\nu}$ and magnetic background $\Sp_\rv$.
In order to include a spin texture made of an AF-SkX phase we perform Monte Carlo simulations with overrelaxation updates\cite{rosales2015three,albarracin2016field} for system sizes of  $N = L^2$ sites ($L = 12-84$) and periodic boundary conditions on the following pure magnetic Hamiltonian \cite{rosales2015three,diaz2019topological} (see Supplemental Material \cite{SupplementalM}, Sec. I, for more details on the simulations).  

\bea
H_S&=&\sum_{\langle\rv,\rv'\rangle}J\,\Sp_{\rv}\cdot\Sp_{\rv'}+\Dp_{\rv\rv'}\cdot(\Sp_{\rv}\times\Sp_{\rv'})-B\sum_{\rv}S^z_{\rv}\qquad
\label{eq:Hm}
\eea

\noindent  where $J$ and $\Dp_{\rv\rv'} = D (\rv' - \rv)/\norm{\rv' - \rv}$ are the  antiferromagnetic and DM nearest neighbour couplings respectively (see Fig.~\ref{fig:lattice}a), and $B$ the strength of the magnetic field along the $z$ axis. We will consider $D/J=0.5$ for the rest of the paper as a representative value \cite{rosales2015three,diaz2019topological} where it is know that from the model in Eq.~(\ref{eq:Hm}) emerges a AF-SkX consisting of the superposition of three interpenetrated ferromagnetic skyrmion lattices \cite{rosales2015three,diaz2019topological}(Fig. (\ref{fig:lattice}c) correspond to a portion of the lattice for $B/J=4.6$. See Supplemental Material \cite{SupplementalM}, Sec. I.

In a recent work it was shown that the mixed dynamics of both electrons and spins tend to stabilize the AF-SkX phase in the adiabatic regime $J_H/t \gg 1$  \cite{reja2020skyrmion} allowing us to fix the magnetic texture in Eq.  (\ref{eq:hamiltonian_tb}) throughout our calculations. With this in mind, let's focus in this regime $J_H/t \gg 1$ where the spin of the electrons are aligned parallel to the local moment and the low-energy physics can be described by an effective Hamiltonian of spinless fermions as \cite{ohgushi2000spin} (see Supplemental Material \cite{SupplementalM}, Sec. II for details).

\bea
\label{eq:hamiltinian_eff}
H_{eff} = \sum_{\rv,\rv'}t^{eff}_{\rv\rv'} \dd_{\rv} \d_{\rv'},
\eea
where $\dd_{\rv}$ ($\d_{\rv}$) is the creation (annihilation) operator, $\cos{\theta_{\rv\rv'}} = \Sp_{\rv} \cdot \Sp_{\rv'}$, $t^{eff}_{\rv\rv'}=t_{\rv\rv'}\cos(\theta_{\rv\rv'}/2)e^{i\,a_{\rv\rv'}}$ is the effective hopping amplitude and the phase $\tan(a_{\rv\rv'})=-\sin(\phi_{\rv}-\phi_{\rv'})/(\cos(\phi_{\rv}-\phi_{\rv'})+\cot(\theta_{\rv}/2)\cot(\theta_{\rv'}/2))$.

The electronic band structure and other quantities of interest are obtained through numerical diagonalization of the resulting Hamiltonian matrix in the reciprocal space. Therefore, once the eigenvector $\ket{u_{n}{\left(\kv\right)}}$ and eigenenergies $\epsilon_{n}{\left(\kv\right)}$ are determined, we calculate the Hall conductivity $\sigma_{xy}$ by means of the standard Kubo formula, which at $T=0$, reduces to 
\bea
\sigma_{xy}&=& \frac{e^2}{2\pi h} \sum_n\int \Theta(\epsilon_n-\epsilon_F)\Omega^{(n)}_{xy}\,d^2k
\eea
\noindent where the Berry curvature for the band $n$ is $\Omega^{(n)}_{ab}=2 \sum_{m\neq n}\Im{\left[v_{a}\right]^{n\,m}\left[v_{b}\right]^{m\,n}}/\left(\epsilon_{m} - \epsilon_{n}\right)^{2}$, $\epsilon_{F}$ is the Fermi energy and $\left[v_{a}\right]^{nm} = \bra{u_{n}} v_{a} \ket{u_{m}}$ $(a= x, y)$ is the matrix element of the velocity operator ${\bf v} = \frac{i}{\hbar}\left[H_{eff}, \Rv\right]$, with the position operator being ${\Rv} = \sum_{\rv} \rv\, d^{\dagger}_{\rv} d_{\rv}$.
When the Fermi energy $\epsilon_{F}$ lies inside a band gap, the Hall conductivity is quantized as $\sigma_{xy}= e^2/h \sum_{n} C_{n}$, where the integers $C_{n}$ are the so-called Chern numbers.
Relevant Chern numbers are calculated independently of $\sigma_{xy}$ through the Fukui-Hatsugai numerical method \cite{fukui2005chern,takahiro2013chern}. Thus, due to the the bulk-boundary correspondence principle \cite{hatsugai1993chern}, in an open-boundary system one would expect to find a number $|\nu|$ of topologically protected chiral edges states crossing the $n$th band gap, where $\nu =\sum_{n} C_{n}$. In addition, we also calculate the out-of-plane component of the orbital magnetization \cite{chang1996berry,xiao2005berry,raoux2015orbital}

\bea
\mathcal{M}^{z} &=&\mathcal{M}_c+\mathcal{M}_t\\
&=&  \frac{1}{(2\pi)^2} \sum_{n} \int_{BZ} \Theta_n\,m^{z}_{n}\,d^2k\nonumber\\
&&+\frac{e}{4\pi h}\sum_{n} \int_{BZ} \Theta_n(\Omega^{(n)}_{xy}-\Omega^{(n)}_{yx})(\epsilon_F-\epsilon_n) d^2k\nonumber
\label{eq:OM}
\eea
\noindent where $\Theta_n=\Theta(\epsilon_n-\epsilon_F)$ is the Heaviside step function, $m^{c}_{n}=-\frac{e}{2\hbar} \sum_{m\neq n}\epsilon^{abc}\Im{\left[v_{a}\right]^{n\,m}\left[v_{b}\right]^{m\,n}}/\left(\epsilon_{m} - \epsilon_{n}\right)$ is the crystal orbital magnetic moment with $\epsilon^{abc}$ is the Levi-Civita tensor (summation over $a,b$ is implicit).  $\mathcal{M}_c$ correspond to the contribution from the intrinsic orbital moment (conventional part), whereas $\mathcal{M}_t$ corresponds to corrections of topological nature.

\begin{figure}[th]
\includegraphics[width=1.0\columnwidth]{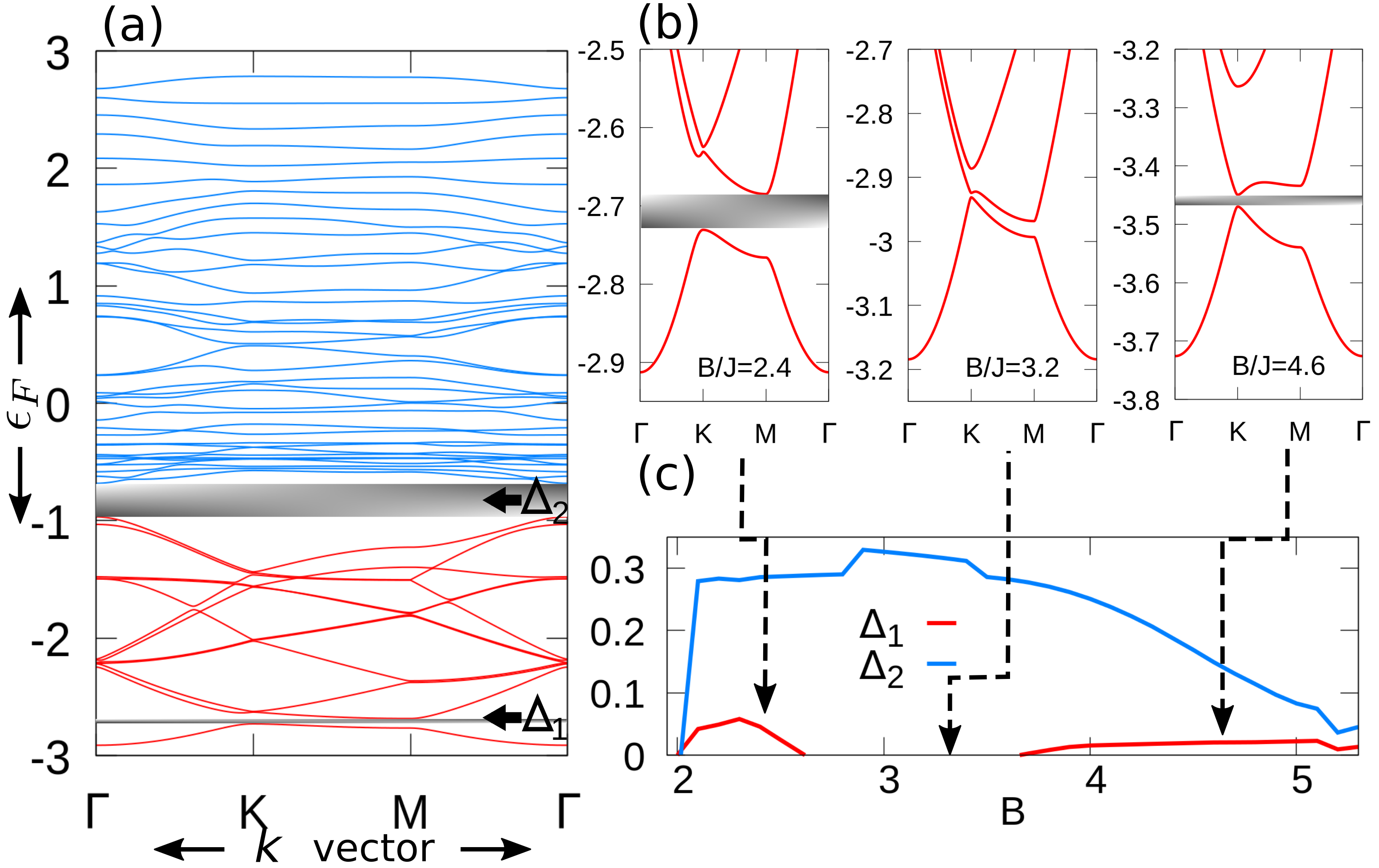}
\caption{(Color online) (a) Representative band structure of the model in Eq.(\ref{eq:hamiltinian_eff}) for $B/J=2.4$ (energy in units of $t$). Red and blue set of curves indicate the two regions with different properties. The gray shaded areas mark the two relevant gaps of the system. (b) Closing and further reopening of the bulk-gap $\Delta_1$. (c) Evolution of the gaps  $\Delta_1$ and $\Delta_2$ vs magnetic field $B$.} 
\label{fig:bands_N_Gaps}
\end{figure}

{\it Band Structure and Topological Phase Transition.--} In Fig. \ref{fig:bands_N_Gaps}a we show a typical band structure obtained in the presence of an AF-SkX background. Here we identify two main distinct energy regions: (i) a low-energy sector consisting of strongly overlapping bands except for a field dependent global bulk gap $\Delta_{1}$ between the first and second band (red curves), and (ii) a high-energy sector resembling a typical THE spectrum in the presence of a ferromagnetic Skyrmion background \cite{gobel2017unconventional} (blue curves). Both sectors are separated by a persistent energy gap $\Delta_{2}$ (see Fig. \ref{fig:lattice}c).

In this section, we focus our attention on the low energy sector, leaving the discussion for the high energy sector to the next section.
For sequentially bigger fields it is found that the overlapping region barely changes.
Fig. \ref{fig:bands_N_Gaps}(b) shows a closing and further reopening of the first bulk gap $\Delta_1$ for increasing fields. These two sectors where $\Delta_1$ is non-zero (at low field and a high field) are well separated by a broad gapless region as shown in Fig. \ref{fig:bands_N_Gaps}(c).
\begin{figure}
\includegraphics[width=0.99\columnwidth]{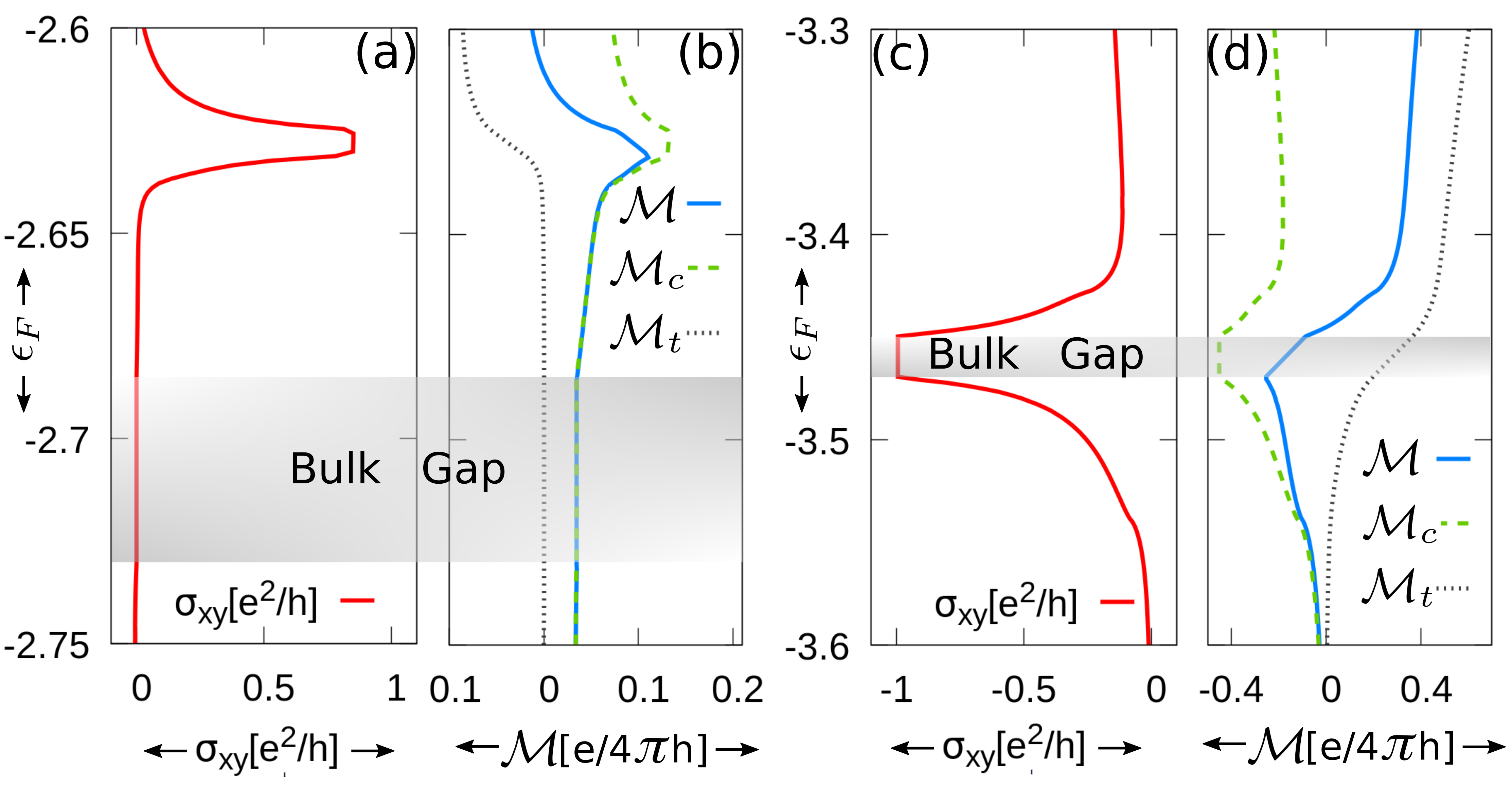}
\caption{(Color online) $\epsilon_F$ dependence (in units of $t$) of the topological Hall conductivity $\sigma_{xy}$ (panels  (a) and (c)), exhibiting a  transition from zero to quantized value in the bulk gap window as a function of Fermi energy.  Panels (b) and (d) show the orbital magnetization $\mathcal{M}$ (and $\mathcal{M}_c$, $\mathcal{M}_t$) in units of $[e/4\pi\,h]$. The shaded areas correspond to the energy gap $\Delta_1$.} 
\label{fig:top_phase_tr}
\end{figure}
Upon further inspection we finds a noticeable change in both the Hall conductivity and orbital magnetization when the Fermi energy sits inside the band gap. In Fig. \ref{fig:top_phase_tr} we show this quantities for low field (left) and high field (right) displaying a transition from $\sigma_{xy}=0$ at low field  to a quantized value $\sigma_{xy}=-\frac{e^2}{h}$ at high field (Fig. \ref{fig:top_phase_tr}a and \ref{fig:top_phase_tr}c). This is confirmed by the calculation of the Chern number where the lowest band is topologically  trivial with $C_1=0$ at low field and non-trivial with $C_1=-1$ at high field. This topological change is also evidenced in the orbital magnetization (Eq. \ref{eq:OM}). In Fig.  \ref{fig:top_phase_tr}(panel b and d) we show the orbital magnetization ($\mathcal{M}$) and its two components $\mathcal{M}_c$ and $\mathcal{M}_t$  as a function of $\epsilon_F$. On one hand, we observe that in the energy gap, at low field there is only a negative contribution from the conventional part $\mathcal{M}_c$. Because $C_1=0$ we have $\mathcal{M}_t\equiv0$. On the other hand, at high field, $\mathcal{M}_c$ keeps constant inside the gap, due to the fact that the integral of  $m^{z}_{n}$ does not depend on $\epsilon_F$. In contrast, the Berry-phase term $\mathcal{M}_c$  linearly increases with $\epsilon_F$, as is expected from Eq. (\ref{eq:OM}). 

\begin{figure}[htb]
\includegraphics[width=0.9\columnwidth]{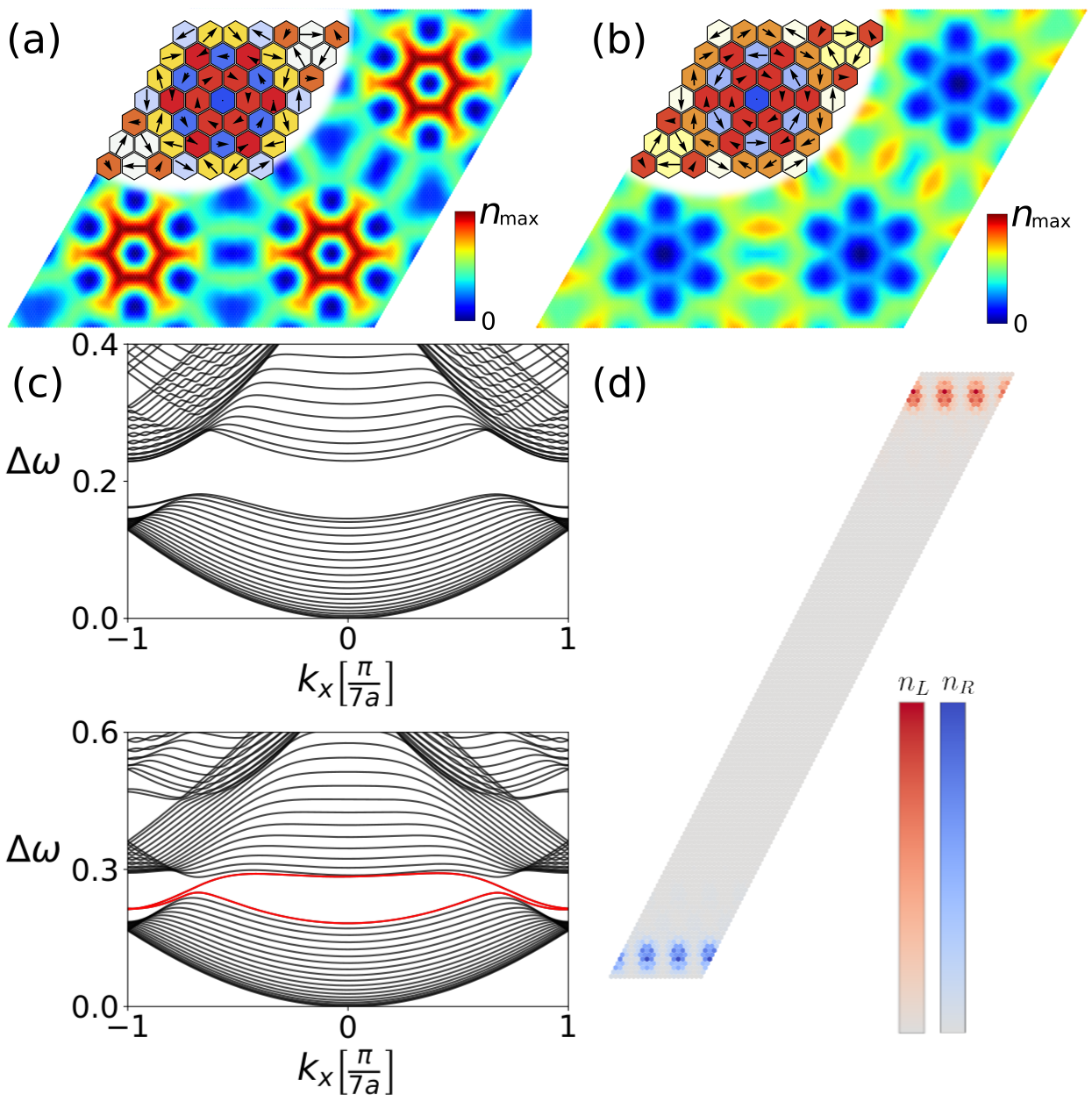}
\caption{(Color online) Electronic occupation at the $\Gamma$ point for the lowest band for $B/J=2.4$ (a) and $B/J=4.6$ (b). In the trivial case ($C_1=0$) the electron density is strongly localized inside the AF-Skyrmion; when $C_1=-1$ it becomes delocalized (the top-left insets show the spin configurations). (c) One dimensional band structure ($\Delta w=\epsilon_n-\epsilon_1$ in units of $t$) for a nanoribbon geometry with periodic (upper) and open (bottom) boundary conditions. In the last case, the edge state (red curve)  is clearly observed between the two adjacent bands, which demonstrates a  nonzero Chern number. (d) The edges states are localized at the boundary (top and bottom) of the nanoribbon.} 
\label{fig:ocup_gamma}
\end{figure}

In order to inspect in more detail the effects of the transition, we calculate the electronic occupation $n_{\rv}=\langle u_1(\kv)|\hat{d}^{\dagger}_{\rv}\,\hat{d}_{\rv}|u_1(\kv) \rangle$ within the unit cell at $\Gamma$ point for the lowest band for $B/J=2.4$ and $4.6$. In the low-field region it is strongly confined around
the anti-ferromagnetic skyrmion center in a ring-like configuration\cite{redies2020mixed} (Fig. \ref{fig:ocup_gamma}a). As we cross the transition the electronic occupation spreads out, getting away from the skyrmions centers, forming a connected configuration throughout the system (Fig. \ref{fig:ocup_gamma}b). This localized/delocalized like transition  is compatible with the change in Hall conductivity. 

Lastly, we study the presence of in-gap chiral edge states. To this end, we have numerically diagonalized the Hamiltonian Eq. (\ref{eq:hamiltinian_eff}) in a one-dimensional strip configuration with a width of $20$ unit cells. The calculated band structure with periodic and open boundary conditions is  shown in Fig. \ref{fig:ocup_gamma}c. In the high-field region, a gap crossing edge state, absent in the low field region, is observed. The electronic occupation for this state localizes in the edge of the sample, as shown in Fig. \ref{fig:ocup_gamma}d. A similar behaviour can be found in the persistent energy gap $\Delta_2$ with stable in-gap chiral edge states (see Supplemental Material \cite{SupplementalM}, Sec. II, Fig. 2). This result agrees with the bulk-boundary correspondence principle.

Therefore, a system of this kind with a filled first band could be tuned from an insulator, to a metal, to a Chern insulator phase by changing the external magnetic field. Edge conducting states can be turned on and off in the same manner.

\begin{figure}[thb]
\includegraphics[width=0.98\columnwidth]{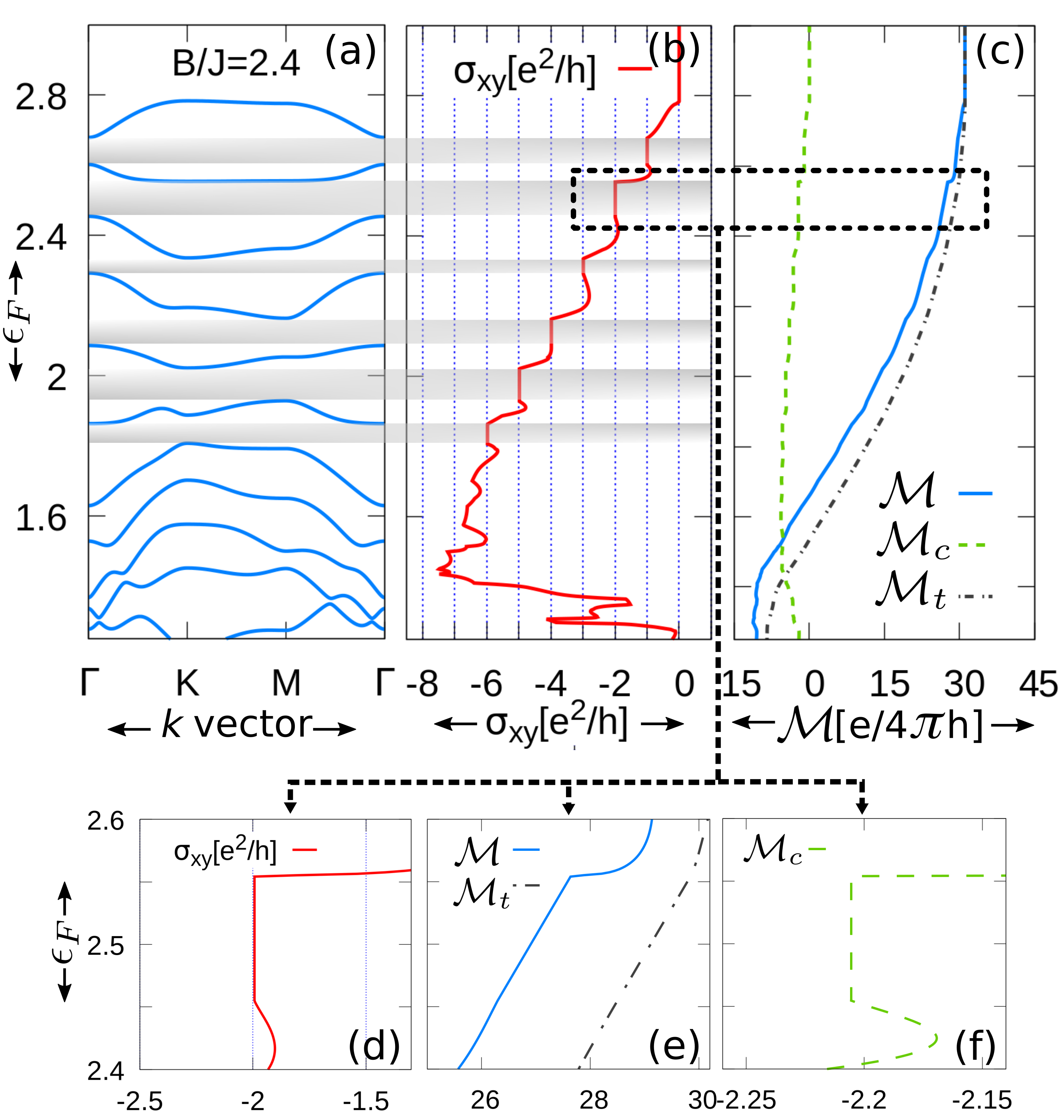}
\caption{(Color online) (a) Electronic band structure at high filling band (in units of $t$), (b) Hall conductivity  and (c) orbital magnetization as a function of the Fermi energy (vertical axis). Regions with bulk gap are indicated by a gray rectangular box. In panels (d),(e) and (f) we show a zoomed region in order to highlight the connection between quantities.} 
\label{fig:edge_states}
\end{figure}

{\it High filling sector.--} At high filling, the band structure, Hall conductivity and orbital magnetization show a striking similarity to that observed in ferromagnetic skyrmion crystals \cite{hamamoto2015quantized, gobel2017unconventional} as function of the Fermi energy (see Fig. \ref{fig:edge_states}). The presence of THE is a consequence of the three sublattice structure of the AF-SkX magnetic background. On an AF-SkX on a bipartite lattice one would expect the emergent field to fluctuate around zero, leading to a vanishing Hall conductivity \cite{gobel2017antiferromagnetic}. However, in the case of the AF-SkX on the triangular lattice, the emergent field fluctuates around a non-zero value and its strength is comparable to that of the emergent field coming from a ferromagnetic Skyrmion lattice background. In this sense, a correspondence between the Integer Quantum Hall Effect and the THE on the AF-SkX could be traced, as is the case of ferromagnetic Skyrmion lattices \cite{gobel2017unconventional}.

%

{\it Conclusions.--} In this letter we have studied the Topological Hall Effect and Orbital Magnetism of electrons coupled to an  antiferromagnetic skyrmion lattice. 
The band structure consists of two energy regions, separated by a persistent energy gap. The low-energy sector consists of a strongly overlapping bands except for a switchable bulk gap between the first and second band. The high-energy sector shows striking similarity to that of the Integer Quantum Hall Effect.

At low-filling,  we found that a magnetic field drives the system from a conventional insulating   state to a topological insulator state hosting chiral edge states in generic strip geometries. 
This topological change is clearly  manifested in the Chern numbers, the electron density, Hall conductivity and orbital magnetization. 
In the region with Chern number $C_1=0$, the electron density is strongly localized at the AF-Skyrmion cores forming a ring-like distribution. In the region with $C_1=-1$, it becomes delocalized. We found that the localization/delocalization of the ring states can be controlled by the magnetic field which determines the texture details. The Hall conductivity presents a switchable behaviour from a null to a  quantized value $\sigma_{xy}=-e^2/h$ when the Fermi energy is inside the bulk gap. We have found that the two parts $\mathcal{M}_c$ and $\mathcal{M}_t$ of the orbital magnetization displays a fully different behaviors in the $C_1=0$ and $C_1=-1$ regions because of their different roles in these two regions.

At high fillings, even having a three sublattice structure of the AF-SkX  we recover a similar behavior observed in a ferromagnetic Skyrmion lattice where it is possible to connect the THE with the Integer Quantum Hall Effect.

Our results highlight the richness of the electronic phases arising from systems hosting AF-SkX states and their potential as platforms for spintronic devices. The present study on the THE in antiferomagnetic skyrmion lattices calls for experimental verification. The low filling topological phase transition and controlled chiral edge states can be studied in materials which exhibit a AF-SkX phase, e.g., the recent fractional skyrmion lattice observed in  the compound MnSc$_2$S$_4$ \cite{ShangGao2020Nat}. In this material, magnetic ions Mn$^{2+}$ form a diamond lattice. At low temperature and finite magnetic field,  each triangular lattice layer along the [111] direction  realizes a fractional AF-SkX that is composed of three ferromagnetic Skyrmion sublattices.

{\it Acknowledgments.--} We thank Pierre Pujol for valuable comments. H.D.R thanks Flavia G\'omez Albarrac\'in for fruitful discussions. This work was partially supported by CONICET (PIP 2015-813), ANPCyT (PICT 2012-1724) and SECyT-UNLP (PPID X039). H.D.R. acknowledges support from PICT 2016-4083. 

\bibliographystyle{apsrev4-1}
\bibliography{refsv3}

\clearpage
\onecolumngrid

\end{document}